\documentstyle[prl,aps]{revtex}
\begin{document}
\twocolumn

\noindent{\bf Reply to Comment on ``Quantum Pump for Spin and Charge
Transport in a Luttinger Liquid''}

In their Comment \cite{1}, Feldman and Gefen dispute our results in
Ref.~\cite{2} by claiming that the high frequency pumping regime that we
employ is peculiar. We disagree with this assessment, and we show below that
their argument revolves around a trivial issue - it is simply a question of
the relative magnitudes for the {\it two} crossover scales in the problem,
$\Omega_c$ and $\omega_\Gamma$. In our paper we analyze the case
$\omega_\Gamma<\Omega_c$, and they discuss $\omega_\Gamma>\Omega_c$.

As we state in our paper, for pumping frequencies $\omega_0<\Omega_c=v_F/a$
($v_F$ is the Fermi velocity and $a$ is the separation between the two
impurities that drive the pump), the two-barrier (or two-impurity) problem
can be written as an effective one-barrier (or one-impurity) problem. Above
this crossover scale $\Omega_c$, the two impurities are resolved \cite{3},
and one cannot use results for the single-impurity problem \cite{4,5}.

The other energy scale is the crossover scale for the effective single
impurity problem, $\omega_\Gamma$, which depends on the backscattering
amplitude $\Gamma$ and the cut-off energy scale $E_F$, $\omega_\Gamma
\sim E_F (|\Gamma|/E_F)^{1/1-g}$.

By focusing on the case $\omega_\Gamma<\Omega_c$, we studied the problem
completely within the single-impurity regime. In this case, the ultraviolet
(UV) limit corresponds to frequencies $\omega_0\gg \omega_\Gamma$, and the
infrared (IR) limit corresponds to $\omega_0\ll \omega_\Gamma$, exactly in
the same sense as in Ref.~\cite{4,5}. For this one-impurity problem, as we
argue in the paper, the UV pumping conductance reaches a universal value.

Feldman and Gefen analyze the case $\omega_\Gamma>\Omega_c$ instead. In this
situation, one does not reach the UV limit of the single-impurity problem,
and, thus, the pumping frequency $\omega_0\approx \Omega_c$ is still in the crossover
region where the pumping conductance is, obviously, non-universal.

We would like to point out, however, that the authors of Ref. \cite{1} choose
incorrectly the high energy cut-off scale in their perturbative calculation.
The ultraviolet cut-off scale should be of order $E_F$, and not $\Omega_c$.
Contrary to what they state, $\omega_\Gamma$ depends on $\Gamma$ and $E_F$,
and not $\Omega_c$.

Having specified the frequency regime we are considering in terms of the two
crossover scales $\Omega_c$ and $\omega_\Gamma$ in the problem, we hope it is
clear what the UV limit of the effective theory means. 

Finally, we would like to comment that Feldman and Gefen's argument for
quantization in the IR limit is interesting and useful in the bosonization
context. We too have recently advanced a more general proof for the
quantization (without using bosonization), also applicable to systems not
described by the Luttinger model with short-range interactions~\cite{6}.\\

\noindent Prashant Sharma and Claudio Chamon
\newline Dept. of Physics, Boston University, Boston, MA 02215



\bigskip

\begin{references}
\frenchspacing
\bibitem{1} D. E. Feldman and Y. Gefen, Comment.
\bibitem{2} P. Sharma and C. Chamon, Phys. Rev. Lett. {\bf 87}, 096401
(2001).
\bibitem{3} C. Chamon, D. E. Freed, S.A. Kivelson, S. L. Sondhi, X. G. Wen,
Phys. Rev. B {\bf 55}, 2331 (1997).
\bibitem{4} C. L. Kane and M. P. A. Fisher, Phys. Rev. B {\bf 46}, 15233
(1992).
\bibitem{5} P. Fendley, A.W.W. Ludwig, H. Saleur, Phys. Rev. B {\bf 52}, 8934
(1995).
\bibitem{6} P. Sharma and C. Chamon (to be published).
\end{references}
\end{document}